\documentclass[prl,twocolumn,superscriptaddress,showpacs]{revtex4} % - FINAL

\usepackage{amsmath,amssymb}
\usepackage{graphicx}
\usepackage{dcolumn}
\usepackage{bm}

\begin{document}

\title{Superfluid--Insulator Transition in Commensurate One-Dimensional Bosonic System with Off-Diagonal Disorder}

\author{Kar\'en G. Balabanyan}%
\affiliation{Department of Physics, University of Massachusetts, MA 01003}%
\author{Nikolay Prokof'ev}%
\affiliation{Department of Physics, University of Massachusetts, MA 01003}%
\affiliation{Russian Research Center ``Kurchatov Institute", 123182 Moscow, Russia}%
\author{Boris Svistunov}%
\affiliation{Department of Physics, University of Massachusetts, MA 01003}%
\affiliation{Russian Research Center ``Kurchatov Institute", 123182 Moscow, Russia}%
\affiliation{Kavli Institute for Theoretical Physics, University of California, Santa Barbara, CA 93106}
\date{\today}% It is always \today, today, but any date may be explicitly specified

\begin{abstract}
We study the nature of the superfluid--insulator quantum phase transition in a one-dimensional system of lattice bosons with off-diagonal disorder in the limit of large integer filling factor. Monte Carlo simulations of two strongly disordered models show that the universality class of the transition in question is the same as that of the superfluid--Mott-insulator transition in a pure system. This result can be explained by disorder self-averaging in the superfluid phase and applicability of the standard quantum hydrodynamic action. We also formulate the necessary conditions which should be satisfied by the stong-randomness universality class, if one exists.
\end{abstract}

 \maketitle
An interplay between commensurability, interactions, and disorder in the superfluid--insulator quantum phase transition is a challenging theoretical problem with applications to such diverse physical systems as $^{4}$He in porous media and aerogels, superfluid films on various substrates, Josephson junction arrays, granular superconductors, disordered magnets, etc. (see, e.g., \cite{Fisher89,Giarmarchi,Herbut,Svist,PS04,Altman}, and references therein). Recently, research on disordered bosons is strongly stimulated by the possibility of controlled experiments with ultracold atomic systems \cite{Damski,Wang}.

Commensurability is relevant for the system of lattice bosons with the off-diagonal disorder (random hopping amplitude or on-site repulsion) and exact particle-hole symmetry (taking place in the limit of large filling factor) \cite{DiagDis}. It leads to a new type of insulator, incompressible and gapless Mott glass (MG) \cite{rem1} in place of incompressible gapped Mott insulator (MI). In two-dimensional (2D) systems, changing the nature of the insulating phase results in the new universality class of the superfluid-insulator (SF-I) transition. The SF-MG transition is characterized by the dynamical critical exponent $1<z<2$ different from  $z=1$ for the SF--MI point in  a perfect system \cite{PS04}. A similar situation is expected in 3D.

Surprisingly, very little is known for fact about the 1D case apart from perturbative renormalization group (RG) arguments. On the superfluid side of the transition, one can use the instanton language \cite{Kashurnikov} in terms of which weak off-diagonal disorder does not seem to be relevant, leading only to the inhomogeneity of the microscopic stiffness in the equivalent (1+1)D classical $XY$ model. Recently,  Altman {\it et al.} \cite{Altman} argued on the basis of the spatial RG analysis that while small off-diagonal disorder is indeed irrelevant for the SF--MG criticality, there should exist also the strong-randomness fixed point. We are not aware of any large-scale numerical simulations attempting to study the SF--MG criticality in the strongly disordered system.

The main observable of interest in 1D systems is the Luttinger-liquid parameter $g=\pi\sqrt{\Lambda\kappa}$, where $\Lambda$ is the superfluid stiffness and $\kappa$ is the compressibility. There is a ``smoking gun" signature in the behavior of $g$ at the critical point which allows to discriminate between different scenarios of the phase transition. If the SF-MG transition is in the same universality class---at least on the superfluid side---as the SF-MI point in a pure system then $g$ should jump discontinuously from $g_c=2$ to zero and obey the Kosterlitz-Thouless RG equations in finite-size systems (the corresponding jumps in $\Lambda$ and $\kappa$ are not universal). The prediction of Ref.~\cite{Altman} for the strong-randomness fixed point is that only $\kappa$ obeys the Kosterlitz-Thouless (KT) scenario \cite{rem2}. Our approach then is to simulate the (1+1)D classical analog of the bosonic Hubbard Hamiltonian using Worm Algorithm \cite{PS04}, and to see whether the KT scenario with the universal jump $g_c=2$ is taking place.

We simulate two microscopically different models of strong off-diagonal disorder. In both cases we observe agreement with the ideal KT scenario. This fact and the observation that $\Lambda$ and $\kappa$ are self-averaging quantities near the transition point allow us to put forward an asymptotic quantum hydrodynamic approach. We argue that instantons of unit charge inevitably become relevant when $g$ is small enough and thus any consistent strong-randomness scenario should necessarily predict discontinuous behavior of all three quantities $\Lambda$, $\kappa$, and $g$.

At large integer filling factor, the bosonic Hubbard model is equivalent to the quantum rotor Hamiltonian (see, e.g., \cite{Sachdev}):
\begin{equation}
H = - \sum_j \, t_j \cos (\varphi_{j+1} - \varphi_j ) + \sum_j \, \frac{U_j}{2} \, \left(\frac{1}{i}\frac{\partial}{\partial \varphi_j} \right)^2
 \; . \label{rotor}
\end{equation}
The first term describes particle hopping between the nearest neighbour sites and the second term describes the on-site repulsion.

While it is possible to perform a direct simulation of the quantum model (\ref{rotor}), we prefer to work with its macroscopic classical equivalent---a version of the bond-current model \cite{Wallin94} described by the anisotropic action:
\begin{equation}
S = {1\over 2} \sum_{\bf n}\, (\tilde{t}_{x} J_{{\bf n},\hat{x}}^2 + \tilde{U}_{x} J_{{\bf n},\hat{\tau}}^2)
 \; . \label{H_bond}
\end{equation}
Here integer vector ${\bf n}=(x,\tau)$ labels sites of the two-dimensional square space-time lattice of linear size $L$, $\hat{x}$ and $\hat{\tau}$ are unit vectors pointing in the space and discrete imaginary time directions, respectively, and $J_{{\bf n},\alpha}$ ($\alpha = \hat{x}, \hat{\tau}$) are integer ``currents" living on lattice bonds. The allowed configurations of bond currents are subject to the zero-divergence constraint $\sum_{\alpha} (J_{{\bf n}, \alpha} + J_{{\bf n}, -\alpha}) = 0$, where, by definition, the direction $-\alpha$ is understood as opposite to $\alpha$ and $J_{{\bf n}, -\alpha}=-J_{{\bf n}-\alpha, \alpha}$.

As its limiting case, the bond-current model (\ref{H_bond}) reproduces the (1+1)D worldline action of the Hamiltonian (\ref{rotor}), the correspondence being given by the relations
\begin{equation}
\tilde{t}=-2\ln \left(\frac{t \! \cdot \! \Delta \tau}{2}\right)
 \; , ~~~~~~~~ \tilde{U}=U \! \cdot \! \Delta \tau
 \; ,\label{rels}
\end{equation}
where $\Delta \tau$ is the imaginary time step and the limit of $\Delta \tau \to 0$ is assumed. The trick, however, is to consider essentially finite $\Delta \tau$; this may not change the universality class of the continuous phase transition but substantially improves the algorithm performance \cite{Wallin94,Alet,PS04}.
\begin{figure}%[!h]
\includegraphics[width=0.95\columnwidth, height=0.75\columnwidth]{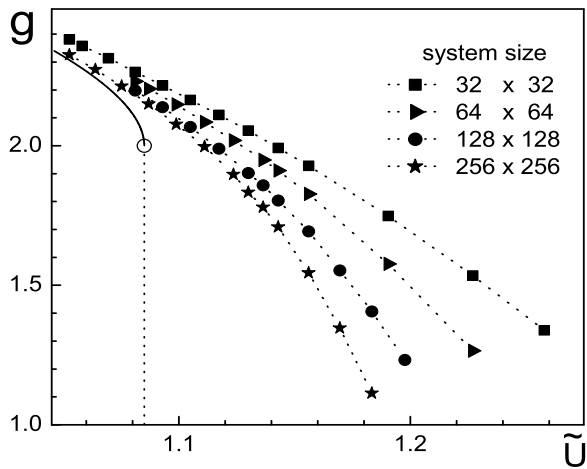}
\caption{\label{fig1}Luttinger-liquid parameter $g$ for the bimodal distribution of $\tilde{t}_{x}$. Statistical errors are smaller than the symbol size. The solid line is the KT extrapolation to the infinite system size. Dotted lines are to guide an eye.}
\end{figure}
\begin{figure}%[!h]
\includegraphics[width=0.95\columnwidth, height=0.75\columnwidth]{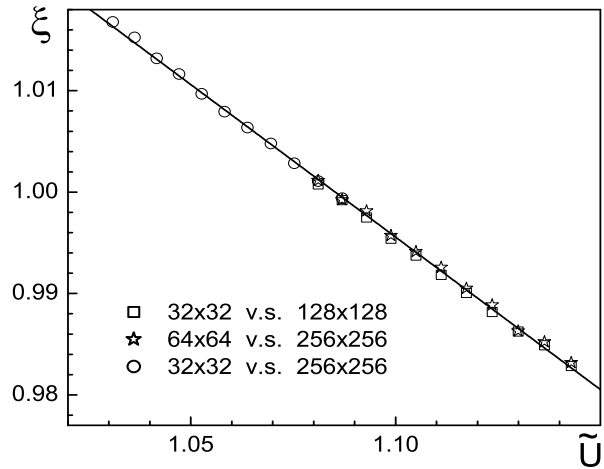}
\caption{\label{fig2}RG parameter $\xi$ for different data sets as a function of control parameter $\tilde{U}$ for the bimodal distribution of $\tilde{t}_{x}$. The solid line is a linear fit.}
\end{figure}

For each system size we considered $2\cdot 10^2\div 2\cdot 10^3$ realizations of disorder. The analysis of sample-to-sample fluctuations shows that there is self-averaging of $\Lambda$ and $\kappa$ near the SF--MG transition for both  models of disorder. Qualitatively, the behavior of $\Lambda$ and $\kappa$ is similar to that of $g$ and indicates the presence of a jump at the critical point. In what follows, we discuss only the Luttinger-liquid parameter $g = \pi\sqrt{\Lambda\kappa}$ because its jump is expected to be universal and can be deduced from the quantitative RG analysis of finite-size effects.

The first set of simulations was performed for fixed $\tilde{U}_{x}\equiv \tilde{U}$ (i.e. no disorder in the on-site interaction) and the bimodal distribution for the hopping parameter $\tilde{t}_{x}$: with equal probabilities we choose $\tilde{t}_{x}=2\tilde{U}$ or $\tilde{t}_{x}=2\tilde{U}/3$. Simulation results and comparison with the RG flow for $g$ as a function of system size $L$ are presented in Figs.~\ref{fig1}, \ref{fig2}. When reduced to its integral form, the KT renormalization flow is given by
\begin{equation}
\int_{g(L_2)/2}^{g(L_1)/2} \frac{dt}{t^{2}(\ln t-\xi)+t}\, =\, 4\cdot \ln\left(\frac{L_{2}}{L_{1}}\right) \; , \label{RG}
\end{equation}
where $\xi$ is size-independent microscopic parameter characterizing the vortex fugacity. The procedure of analyzing the data is as follows. For different pairs of system sizes, $L_1$ and $L_2$, and fixed value of $\tilde{U}$ close to the critical point, one solves Eq.~(\ref{RG}) for $\xi$, and verifies that for large system sizes $\xi$ is $L$-independent. Moreover, the $\xi (\tilde{U} )$ dependence must be analytic and thus well approximated by a straight line in the vicinity of the critical point $\tilde{U}=\tilde{U}_c$. In contrast, the $g_L(\tilde{U})$ curves noticeably deviate from the straight line even in a rather small interval around $\tilde{U}_c$, as they develop a square-root cusp and the universal jump at $\tilde{U}=\tilde{U}_c$ in the limit of $L\to \infty$. These features are clearly seen in Figs.~\ref{fig1}, \ref{fig2}.

It seems that the bimodal distribution with the ratio of hopping amplitudes as large as three may be regarded as a strong disorder case since our system sizes accomodate many local disorder fluctuations with $3\div 5$ consequtive bonds with small/large $\tilde{t}$. Still, one may not exclude the possibility that weak links with power law, rather then exponential, distribution function of small $\tilde{t}$ is necessary to realize the strong-randomness fixed point. We thus have studied a system with continuous distribution of links between the nearest neighbour sites. The quantum-rotor prototype for our model was the Hamiltonian (\ref{rotor}) with the $j$-independent $U_j \equiv U$, and hopping amplitudes $t_j$ independently distributed on the interval $(0, t_{\rm max})$ with the power-law probability density $p(t_j) \propto t_j^{\gamma}$, \cite{powerdist}. To emphasize a possible special role of having more frequent weak links, we kept the ratio $t_{\rm max}/U$ constant and used $\gamma$ as a control parameter. The closer is $\gamma$ to zero, the stronger is the effect of weak links. Accordingly, for the bond-current model (\ref{H_bond}) we set $\tilde{U}_x=0.5$ and $\tilde{t}_x=-2\ln( \zeta /2)$, where $\zeta$ is a random number distributed with the probability density $p(\zeta)\propto \zeta^{\gamma}$ on the $(0, 1.5)$ interval.

\begin{figure}%[!h]
\includegraphics[width=0.95\columnwidth, height=0.75\columnwidth]{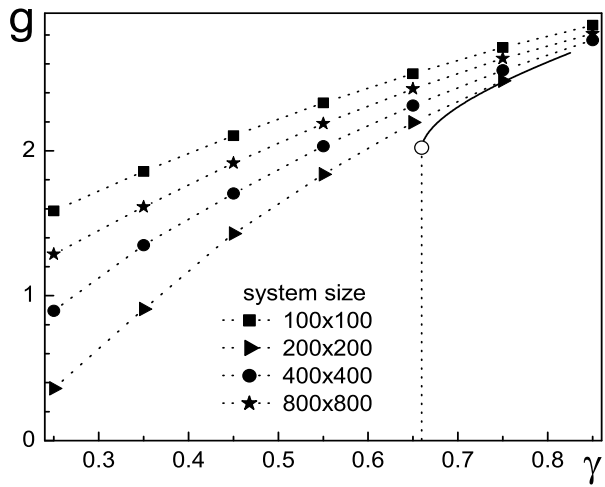}
\caption{\label{fig3} Luttinger-liquid parameter $g$ for the power-law distributed $t_{j}$. Statistical errors are smaller than the symbol size. The solid line is the KT extrapolation to the infinite system size. Dotted lines are to guide an eye.}
\end{figure}

\begin{figure}%[!h]
\includegraphics[width=0.95\columnwidth, height=0.75\columnwidth]{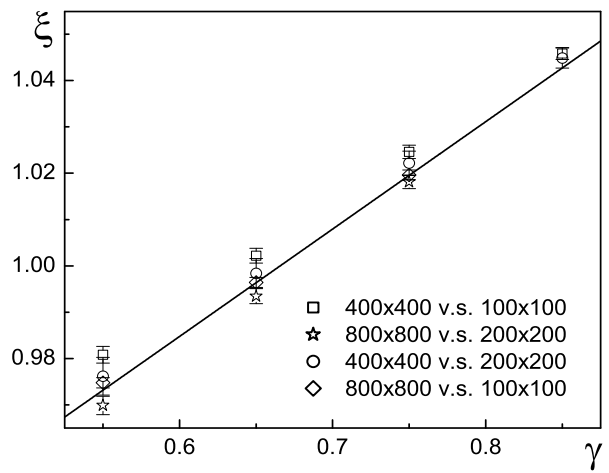}
\caption{\label{fig4}RG parameter $\xi$ for different pairs of data sets as a function of $\gamma$ for the power-law distributed $t_{j}$. The solid line is a linear fit.}
\end{figure}

In Fig.~\ref{fig3} we present our data for $g$ as a function of $\gamma$, which is qualitatively similar to that in Fig.~\ref{fig1}. Note that strong finite-size effects set in only at $g<2$. Also, we observe that self-averaging of $g$ requires very large system sizes $L > 100$ in the vicinity of the transition point. This explains why fitting the data with Eq.~(\ref{RG}), see Fig.~\ref{fig4}, demonstrates stronger finite-size corrections than in the previous case. The overall behavior is consistent with the pure-system KT scaling.


Self-averaging of mesoscopic $\Lambda$ and $\kappa$ was observed in both simulations close to transition point for $g \ge 2$. On the superfluid side of the transition, this outcome is expected and can be proved under two quite general assumptions. If one takes two 1D systems of size $L$ (label them as systems $1$ and $2$) and combines them together to form a new system of size $2L$, then $\kappa (2L)$ and $\Lambda (2L)$ are given by
\begin{align}
\kappa (2L) &= [\kappa_1 (L) + \kappa_2 (L)]/2 \; ,\label{kappa_2L}\\
\Lambda^{-1} (2L) &= [\Lambda^{-1}_1 (L) + \Lambda^{-1}_2 (L)]/2 \; . \label{Lambda_2L}
\end{align}
Equations~(\ref{kappa_2L}) and (\ref{Lambda_2L}) imply that properties of the new system are independent of the junction properties. The latter assumption can hardly be questioned if initial $\kappa$'s and $\Lambda$'s are finite and $L$ is large enough. The self-averaging of $\kappa$ is then guaranteed by the Central Limit Theorem (CLT). The case of $\Lambda$ is less obvious, since the distribution function $f(L,\Lambda )$ may be such that $\langle \Lambda^{-1} \rangle = \infty$, in which case the CTL does not apply. However, if the infinite-size limit for the distribution function is well defined, $f(L \to \infty ,\Lambda ) =f(\Lambda )$, then it must satisfy the following relation
\begin{equation}
f(\Lambda) = \int \delta \left( \Lambda - {2 \Lambda_1 \Lambda_2 \over \Lambda_1 + \Lambda_2}\right)\, f(\Lambda_1) f(\Lambda_2) \, d\Lambda_1 d\Lambda_2 \; . \label{relation}
\end{equation}
Now it is easy to prove that the only solution of Eq.~(\ref{relation}) is the self-averaging one, $f(\Lambda)=\delta(\Lambda - \Lambda_0)$. Indeed, Eq.~(\ref{relation}) leads to the inequality $\langle \Lambda ^2 \rangle \leq \langle \Lambda \rangle^2$, immediately implying that the distribution is dispersionless.

When the self-averaging of $\Lambda$ and $\kappa$ takes place, the long-range groundstate properties of the superfluid phase can be described by the Popov's hydrodynamic action \cite{Popov} with the density-field fluctuations integrated out:
\begin{equation}
S[\Phi] = \int dxd\tau \left\{ { \Lambda\over 2} (\Phi_x')^2 + { \kappa \over 2} (\Phi_\tau')^2 + i n_0(x) \Phi_\tau' \right\} .\label {a}
\end{equation}
Here $\Phi (x,\tau)$ is the phase field, and $n_0$ is the equilibrium value of the local number density. The last term in $S$ is sensitive only to instantons---vortex-type topological defects in the non-single-valued phase field $\Phi (x,\tau)$. Integrating the last term over $\tau$ and introducing a rescaled time variable $y=c\tau$ ($c=\sqrt{\Lambda/ \kappa}$ is the sound velocity), one gets the effective action in the form (see, e.g., \cite{Kashurnikov})
\begin{equation}
S[\Phi] = {g\over 2\pi } \int (\nabla \Phi)^2dxdy + i\sum_\nu \, p_\nu \, \theta(x_\nu) \; ,\label{a2}
\end{equation}
where $\nu$ enumerates the instantons, $p_\nu = \pm 1,\pm 2, \ldots$ and $x_\nu$ are the charge and the coordinate of the instanton $\nu$, respectively, and
\begin{equation}
\theta(x)= -2\pi \int _0^x n_0(x') \, dx' \; . \label{gamma}
\end{equation}
The first term in Eq.~(\ref{a2}) corresponds to the 2D $XY$ model, and the second term describes the position-dependent instanton phase. For a system in a periodic external potential, one can ignore the phase term if it changes by a multiple of $2\pi$ when the instanton is shifted by one lattice period, $a$, i.e. when for any integer $m$
\begin{equation}
\int _0^{ma} \! n_0(x) \, dx \, =\, \mbox{integer} \; . \label{cond}
\end{equation}
This relation is satisfied for a pure system at integer filling factor, but typically does not hold true in the presence of disorder because $n_0(x)$ becomes a random function of $x$. An exceptional situation occurs in a system with off-diagonal disorder and at large integer filling. The particle-hole symmetry is then preserved locally, and the occupation number for {\it each} site $x$ remains integer. Hence, the second term in (\ref{a2}) is irrelevant and the macroscopic behavior of the system is indistinguishable from that of a pure system, i.e. the SF-MG transition is in the 2D $XY$ universality class.

Under these circumstances, one can make the following statement about alternative (to the pure-system one) scenario of the SF--MG criticality. It is possible only if at the critical point it leads to a jump of $g$ with the amplitude greater or equal than $2$. Moreover, at $g_c$ the system should be non-self-averaging, to exclude the applicability of the hydrodynamic approach. Any theory that predicts a SF phase with finite $\Lambda$ and $\kappa$ and $g<2$ is inconsistent, since, in view of the established self-averaging, the hydrodynamic+instanton approach is applicable and superfluidity is destroyed by the KT scenario at $g=2$.

It is worth noting that while the hydrodynamic action (\ref{a2}) is applicable to the SF phase and the critical point, it can not be used on the insulating side of the transition. Formally, for $g<2$ Eq.~(\ref{a2}) predicts the MI phase with the gap in the energy spectrum while the MG phase is gapless (the absence of the gap is due to arbitrary large and thus exponentially rare superfluid regions \cite{PS04}).

%----------------------------------------------------------------------
To conclude, we have presented numerical evidence and general arguments that in a 1D system of lattice bosons with strong off-diagonal disorder the criticality of the SF-MG transition is the same as for the SF-MI transition in a pure system. We did not find the evidence in favor of the strong-randomness fixed point \cite{Altman}. We argued, that an alternative scenario, if in principle possible, is inconsistent with continuously vanishing superfluid stiffness or/and compressibility at the critical point, and must predict a jump of the Luttinger-liquid parameter with the amplitude larger or equal to 2.

We are grateful to E.~Altman, A.~Polkovnikov and G.~Refael for stimulating discussions. The research was supported by the National Science Foundation under Grant No. PHY-0426881, and, in part, under Grant No. PHY99-07949.


\begin{thebibliography}{99}
\bibitem{DiagDis} It was argued on general grounds \cite{Fisher89} and demonstrated in Monte Carlo simulations \cite{PS04} that the diagonal disorder (random on-site potential), no matter how weak, always leads to the superfluid--insulator transition from the superfluid to a compressible gapless Bose glass, rather than to incompressible gapped Mott insulator.
\bibitem{rem1} Apparently, the term Mott glass for a gapless incompressible commensurate disordered insulator has been coined in Ref.~\cite{Giarmarchi}.
\bibitem{rem2} Unfortunately, Ref.~\cite{Altman} does not present a specific prediction for the behavior of $\Lambda$ and $g$ at the strong-randomness critical point.
However, if their behavior turns out to be discontinuous with $g_c=2$, then the difference between two scenarios becomes illusive.
\bibitem{powerdist} The power-law distribution of (renormalized) disorder was argued by the authors of Ref.~\cite{Altman} to be characteristic of the vicinity of the strong-randomness critical point, with $\gamma \to 0$ at the transition point. Hence the power-law disorder at the microscopic level seems to be most natural for achieving the strong-randomness scenario, provided the latter exists.
\end{thebibliography}
\end{document}